\documentclass[aps,prb,preprint,groupedaddress]{revtex4}

\newif\ifpdf
\ifx\pdfoutput\undefined
\pdffalse 
\else
\pdfoutput=1 
\pdftrue
\fi

\ifpdf
\usepackage[pdftex]{graphicx}
\else
\usepackage{graphicx}
\fi

\begin{document}

\ifpdf
\DeclareGraphicsExtensions{.pdf, .jpg}
\else
\DeclareGraphicsExtensions{.ps, .jpg}
\fi

\def\hslash{\hbar}
\def\imag{i}
\def\grad{\vec{\nabla}}
\def\div{\vec{\nabla}\cdot}
\def\curl{\vec{\nabla}\times}
\def\DDt{\frac{d}{dt}}
\def\ddt{\frac{\partial}{\partial t}}
\def\ddx{\frac{\partial}{\partial x}}
\def\ddy{\frac{\partial}{\partial y}}
\def\lap{\nabla^{2}}
\def\divv{\vec{\nabla}\cdot\vec{v}}
\def\gradS{\vec{\nabla}S}
\def\vvec{\vec{v}}
\def\wc{\omega_{c}}
\def\<{\langle}
\def\>{\rangle}
\def\Tr{{\rm Tr}}
\def\Csch{{\rm csch}}
\def\Coth{{\rm coth}}
\def\Tanh{{\rm tanh}}
\def\g2{g^{(2)}}


\title{Interplay between the repulsive and attractive interaction and 
the spacial dimensionality of an excess electron in a simple fluid}

\author{Ashok Sethia}

\affiliation{Department of Chemistry and Center for Materials Chemistry, 
University of Houston, Houston, TX 77204}

\author{Eric R. Bittner}
\affiliation{Department of Chemistry and Center for Materials Chemistry, 
University of Houston, Houston, TX 77204}

\author{Fumio Hirata}
\affiliation{Department of Theoretical Studies, Institute for Molecular Science,
Myodaiji,
Okazaki 444-8585, Japan}

\date{\today}

\begin{abstract}
The behavior of an excess electron in a one, two and three 
dimensional classical liquid has been studied with the aid of Chandler, Singh 
and Richardson (CSR) theory [J. Chem. Phys. {\bf 81} 1975 (1984)] . 
The size or  dispersion of the wavepacket associated 
with the solvated electron is very 
sensitive  to the interaction
between the electron and fluid atoms, and exhibits complicated behavior in
its density dependence. The behavior is interpreted in terms of an interplay
among four causes: the excluded volume effect due to solvent, the pair
attractive interaction between the electron and a solvent atom, the 
thermal wavelength of the electron ($\lambda_e$),  a
balance of the attractive interactions from different solvent atoms and 
the range of repulsive interaction between electron and solvent atom. Electron 
self-trapping behavior in all the dimensions has been studied for the same 
solvent-solvent and electron-solvent interaction potential and the results 
are presented for the same parameter in every dimension to show the 
comparison between the various dimensions.

\end{abstract}


\maketitle


\section{Introduction}

The behavior of an excess eletron in a wide variety of fluids
has been an interesting topic for many years \cite{ref1,ref2,ref3,ref4,ref5,ref6,ref6,ref8,ref9,ref10}
In the gas phase or the dilute liquid phase,
the electron behaves almost like a free particle. As the solvent density
increases, the electron exihibits different properties depending upon the 
nature
of the solvent and electron-solvent interactions. At liquid density, the
electron
may become self-trapped in a cavity of solvent particles or
remain {\it quasifree} depending upon the nature of electron-solvent
interaction.
Observed properties such as the electron mobility\cite{ref11}
and the absorption
spectra \cite{ref12}
probe the nature of the electronic states in the 
fluid and phenomena of localization\cite{ref13}.

There are a broad range of 
problems in condensed matter physics that are intimately related to the 
problem of excess 
electrons in deformable medium. These include charge transfer kinetics 
in biological reactions, metal-insulator transitions in fluids, polarons, 
phonon-assisted hopping of charge carriers in semiconductors and insulators, 
quantum-tunneling, etc. 
While the excess electron problem belongs to the general problem of electrons
in disordered materials, the liquid environment is in many ways different from 
the solid medium. In liquids, the constituent particles can diffuse, and local 
environment around the solute electron can be substantially different from 
that in solid.   

  When an electron is solvated in a polar liquid such as 
water or ammonia, the 
strong anisotropic electron-solvent interaction causes significant local 
modification  of the equilibrium fluid structure \cite{ref4,ref10,ref14}. The electron 
becomes localized in a small 
cavity because molecules in  a solvation shell orient to create a potential 
minimum. Even simple fluids are found to exhibit electron mobilities that 
change by many orders of magnitude as the density of the fluid is altered 
slightly. In super critical helium, for example, the electron mobility drops 
by over 4 orders of magnitude as fluid density increased by a factor of 2 
in low density regime\cite{ref2}. The reason for 
this behavior is strong repulsion between electron and solvent atom. This 
causes the the depletion of the solvent atoms from the region of the electron 
and forms a highly localized state of the electron. 

  In many other nonpolar fluids such as Ar, CH$_4$, etc. the electron 
always remains in a state of high mobility\cite{ref15} comparable to many 
semiconducting materials. An interesting density 
dependence of the mobility has been observed in them. It shows a 
minimum near  critical fluid density and a maximum at liquid density.

  Electronic states in reduced dimension are of considerable
interest. For example, for a system less than two spatial dimension, 
electrons are localized with an infinitesimal amount of disorder \cite{ref16}. 
The interest in problems of electron or phonon propagation in a one 
dimensional random potential stems from the discovery and extensive 
experimental study 
of a certain class of organic or metallo-organic materials \cite{ref17}. These 
matarials exhibit strongly anisotropic, 
quasi-one-dimensional behavior attributed to the fact that they consist of 
long chains, weakly interacting with each other. In many of these, the 
presense of a random potential has been proposed in order to explain their 
behavior. Electronic surface states play an important role in a wide variety 
of physical problems. For example, surface electrons on liquid helium 
has shown many interesting properties 
and led to important theoretical advances such as the spectrum of bound 
electronic states, electron transport on the He surface, effects due to 
deformation of the He surface, and the possibility that the electrons may 
form a two-dimensional crystal\cite{ref18} in the field of low-dimensionality 
physics.   

  The theory for the excess electrons in fluids developed by 
Chandler, Singh, and Richardson\cite{ref1} (CSR) is 
based on the path integral formulation of
quantum theory which maps the behavior of the electron on to that of a 
classical isomorphic polymer\cite{ref19}. The solvent-induced potential surface 
for the self-interaction of the isomorphic polymer is evaluated using an 
integral equation (e.g.,
reference interaction site model). 
With known potential surface, the polymer statistics is solved using 
variational
approach\cite{ref20} that allows the determination of electronic properties and the
structure of the liquid near the electron. The input of the theory is the 
pure solvent structure factor and the electron-solvent particle interaction
potential. The CSR theory in its formulation is applicable to an adiabatic
solvent ({\it i.e., solvent particles are treated classically}), but has been
extended to treat the effect of the quantum mechanical charge density
fluctuation in the solvent particles \cite{ref21}. The calculated 
electron-absorption 
line shape and mobility are in good agreement with the simulations\cite{22} and 
experiments \cite{ref6}. The predictions of CSR theory 
were verified by computer simulations \cite{ref23,ref24}. 

  For a one-dimensional system we \cite{ref25} have shown recently that the 
repulsive and attractive parts of electron-solvent interaction potential 
lead separately to localization of electron, respectively, by creating cavity 
or forming a cluster of the fluids atoms around it. In two dimensional system 
we \cite{ref26} have shown that dispersion of the wavepacket associated with 
the solvated 
electron is very sensitive to the interaction between the electron and the 
fluid atoms, and exhibits complicated behavior in its density dependence. CSR 
theory has been extended to calculate the effective mass \cite{ref27} as well as 
the density matrix of the excess electron in fluid \cite{ref28}.

The CSR theory involves three or more characteristic lengths, depending upon 
the nature of electron-solvent interaction. These lengths are the 
thermal wavelength 
of excess electron $\lambda_e = ({\beta \hbar^2 \over m})^{{1 \over 2}}$ 
(where $\beta$ is the inverse of temperature in the unit of the 
Boltzmann constant $k_B$ , $m$ is the mass of a bare electron, and $\hbar$ 
is the Planck's constant 
divided by 2$\pi$), characteristic length associated with the electron-solvent 
pseudopotential, and a length associated with the mean volume occupied by each 
solvent atom, which is related to $\rho^{\ast^{-{1\over D}}}$, where 
$\rho^\ast 
= \rho_s\sigma^D, \rho_s$ being the number density of the solvent and $\sigma$ 
is the diameter of the solvent atom, and $D$ is the spatial dimensionality of 
the 
system. The behavior of the excess electron is expected to depend sensitively 
on these lengths. Laria and Chandler \cite{ref29} have attempted to explain the 
contrasting behaviors of the electron in super critical helium and xenon on 
the basis of different ranges of the electron-solvent repulsive interactions. 
 
  In the present work we examine in detail the role played by 
different lengths and the spatial dimensionality of the system (in which we 
have considered the same solvent-solvent and electron-solvent model potential)
 to study the 
self-trapping behavior of the electron. The organization of the rest of the 
paper is as
follows. In  Sec. II we briefly review the CSR theory. In Sec. III we have 
presented the results and their discussions. Finally, Section IV presents 
concluding remarks. Appendix A provide some mathematical material for $D$-
dimensional integration 

\section{Theory}

  The system we consider is a single electron dissolved in a 
single component classical solvent. In the CSR theory, an excess electron is
mapped, using a discretized version of the path integral formulation of
quantum mechanics, onto a polymer of P interaction sites or beads. Under this 
isomorphism \cite{ref19}, the electron
can be viewed as a classical ring polymer.

  The total potential energy can be written as 
\begin{eqnarray}
U= U_{es}({\bf r}, \{{\bf R}_i\}) +  U_{ss}(\{{\bf R}_i\}) 
\end{eqnarray}
with 
\begin{eqnarray}
U_{ss}(\{{\bf R}_i\}) = \sum_{i > j =1}^N u_{ss}(|{\bf R}_i - 
{\bf R}_j|),
\end{eqnarray}
and 
\begin{eqnarray}
U_{es}({\bf r}, \{{\bf R}_i\}) = \sum _{i =1}^N u_{es}(|{\bf r} - {\bf R}_i |)
\end{eqnarray}
  Here, {\bf r} denotes the position of the excess electron, 
${\bf R}_i$ is the collection of the coordinates for a solvent atom, and 
N is the number of solvent atoms and $u_{es}(r)$ and $u_{ss}(r)$ are, 
respectively, electron-solvent atom and solvent atom-solvent atom 
interaction potentian. 
We consider a $D$-dimensional system of spheres of diameter 
$\sigma$ in which the pair interaction between the solvent atoms is taken 
to be 
\begin{eqnarray}
u_{ss}(|{\bf r}|)= \cases{\infty&for $|{\bf r}| \le \sigma$;\cr
      {0}&for $|{\bf r}| > \sigma$.\cr}
\end{eqnarray}
where $|{\bf r}|$ is the $D$-dimensional distance (for notational convenience 
the $D$ dependence will not always be explicitly indicated). 
  The electron-solvent atom interaction in a real system consists of a
strong repulsion at short distance due to
orthogonality requirements between
wavefunctions of core electrons in the solvent particle and that of the excess 
electron and attraction
at large distances due to dispersion interaction. However, in a system
of neutral atoms the electronic states are determined primarily by the short
range repulsive interaction or excluded-volume effect. The attractive
interaction becomes important only at low densities. The interaction
between the
electron and solvent atom is taken to be
\begin{eqnarray}
u_{es}(|{\bf r}|)= \cases{ \infty&for $r \leq d$;\cr
      -\epsilon{\exp(-\alpha r)\over \alpha r}&for $r > d$.\cr}
\end{eqnarray}
  Here, $d$ is the distance of closest approach between 
electron and solvent atom.

In CSR theory\cite{ref1} the partition function Z for an electron in a 
bath of classical particles is written as the functional integral
\begin{eqnarray}
Z = \int D{\bf r}(u) \int d\{{\bf R}_i\} \exp\biggl[-{1\over \hbar}\int_0
^{\beta\hbar} du \bigl\{{1\over 2}m|\dot{\bf r}(u)|^2 + 
U_{es} ({\bf r},\{{\bf R}_i\})\bigr\} - \beta U_{ss}({{\bf R}_i})\biggr]
\end{eqnarray}
  where {\bf r}(u) is the electronic path in imaginary time which is 
periodic in time interval $0 \le  u \le \beta\hbar $, i.e., 
{\bf r}(0)={\bf r}($\beta\hbar$). To concentrate our attention on the 
electron degrees of freedom, the partition function given by Eq.(6) can be 
written as
\begin{eqnarray}
Z = Z_s \int D{\bf r}(u) \exp\bigl\{ -\beta S_\circ[{\bf r}(u) - 
\beta\Delta\mu[{\bf r}(u)]\bigr\}
\end{eqnarray}
  where $Z_s$ denotes the partition function of the solvent, 
$\Delta\mu[{\bf r}(u)]$ is the excess chemical potential for the fixed 
electronic path,
\begin{eqnarray}
\hbox{with},\quad\beta S_\circ[{\bf r}(u)] = {1\over \hbar}\int_0^{\beta\hbar} du 
{1\over 2}m|\dot{\bf r}(u)|^2
\end{eqnarray}
and $\exp\{ -\beta\Delta\mu[{\bf r}(u)]\}$ is called the influence functional 
which represents the solvent effects on the electon. In the continuum limit
\cite{ref1},
\begin{eqnarray}
-\Delta\mu[{\bf r}(u)] = \rho_s\hat{c}_{es}(0) + {1\over 2}(\beta\hbar)^{-2} 
\int_0^{\beta\hbar} du\int_0^{\beta\hbar} du^\prime 
v(|{\bf r}(u) - {\bf r}(u^\prime)|), 
\end{eqnarray}
  where $\hat c_{es}(0)$ is the k = 0 spatial Fourier transform 
of $c_{es}(r)$.
\begin{eqnarray}
v(|{\bf r}(u) - {\bf r}(u^\prime)| = -\int d{\bf r}^\prime 
\int d{\bf r}^{\prime\prime} c_{es}({\bf r}^\prime,u) 
\chi_{ss}(|{\bf r}^\prime - {\bf r}^{\prime\prime}|) 
c_{es}(|{\bf r}^{\prime\prime} - {\bf r}|,u^\prime)
\end{eqnarray}
  Here {\bf r} and $u$ appear as independent coordinates, {\bf r} is the 
distance between two sites and u measures the length along the contour of the 
polymer, and
\begin{eqnarray}
\chi_{ss}(|{\bf r} - {\bf r}^\prime|) = 
\langle \delta\rho_s({\bf r}) \delta\rho_s({\bf r}^\prime)\rangle
\end{eqnarray}
is the density-density correlation function of the unpurturbed bath. In 
Eqs. (9) and (10), $c_{es}$ is the direct correlation function. Its value 
is determined from the equation \cite{ref1}
\begin{eqnarray}
\rho_s h({\bf r}) = \int d{\bf r}^{\prime}\int d{\bf r}^{\prime\prime} 
\omega ( |{\bf r} - {\bf r}^\prime| ) c_{es}( |{\bf r}^\prime - 
{\bf r}^{\prime\prime} |)
\chi_{ss} ({\bf r}^{\prime\prime})
\end{eqnarray}
where
\begin{eqnarray}
\omega(|{\bf r} - {\bf r}^\prime|) = (\beta\hbar)^{-1}\int 
d(u-u^\prime) \omega(|{\bf r} - {\bf r}^\prime|; u-u^\prime). 
\end{eqnarray}
is the intrapolymer correlation function.

  Eq. (12) is solved for $c_{es}$ and $h$ using suitable 
closure relation \cite{ref25}. Since all sites of a ring polymer on the average 
are equivalent, the site dependence disappears from Eq.(12) and only the 
zero-frequency component $\omega(|{\bf r}|)$ of the equilibrium response 
function is required in Eq.(12).

  To complete the evaluation of excess chemical potential, the 
electronic path integral still has to be performed. Following Feynmam \cite{ref20} 
and Chandler {\it et. al.}, \cite{ref1} the excess chemical potential for the 
fixed electronic path is mimicked by a Gaussian functional,
\begin{eqnarray}
-\beta\Delta\mu_{\rm {ref}}[{\bf r}(u)] = -\Gamma_\circ + 
{1\over 2}(\beta\hbar)^{-2} 
\int_0^{\beta\hbar} du \int_0^{\beta\hbar} du^\prime  \Gamma(u-u^\prime) 
\times |{\bf r}(u) - {\bf r}(u^\prime)|^2
\end{eqnarray}
 where $\Gamma(u-u^\prime)$ is a solvent-induced force constant 
between different sites on the electron polymer and $\Gamma_\circ$ merely 
determines the zero of energy. The  Bogoliubov inequality provides a upper 
bound for the excess chemical potential,
\begin{eqnarray}
\Delta\mu \leq -\beta^{-1} \ln Z_{\rm {ref}} + <\Delta\mu[{\bf r}(u)] - 
\Delta\mu_{\rm {ref}}[{\bf r}(u)]>_{\rm {ref}}
\end{eqnarray}
Here, $Z_{\rm {ref}}$ is the electronic partion function for the Gaussian 
reference 
system and $<->_{\rm {ref}}$ means the average over the reference system 
weight determined by $S_\circ + \Delta\mu[{\bf r}(u)]$. Minimizing the right 
hand side of Eq.(14) provides the optimal Gaussian reference system.
This procedure leads to the following equations. The correlation function for 
the intrapolymer correlation in k-space:
\begin{eqnarray}
\hat\omega(k,\tau) = \exp[-\frac{k^2R^2}{2D}]
\end{eqnarray}
  where
\begin{eqnarray}
R^2(\tau) = \langle | {\bf r}(u) - {\bf r}(u^\prime)|^2\rangle 
= 4D\sum_{n\geq 1}^\infty A_n[1-\cos(\Omega_n\tau)]
\end{eqnarray}
is the mean square displacement between two points on the electron path 
separated by a imaginary time increament $0\leq u-u^\prime\leq\beta\hbar$
with
\begin{eqnarray}  
A_n = (\beta m\Omega_n^2 + \gamma_n)^{-1} 
\end{eqnarray}
where $\Omega_n = \frac{2\pi n}{\beta\hbar}$,
and
\begin{eqnarray}
\gamma_n = -(D\beta\hbar)^{-1}\int_0^{\beta\hbar} du [1-\cos(\Omega_nu)]
\int {d^Dk k^2\over (2\pi)^D}  v(k) \exp (-k^2R^2(u)/2D).
\end{eqnarray}

We solve Eq.(12) for $h$ and $c_{es}$ using closure relation
\begin{eqnarray}
g = 0 \quad \hbox{for}\quad r \leq d      
\end{eqnarray}
\begin{eqnarray}
c_{es} = -\beta\epsilon{\exp(-\alpha r)\over \alpha r}\quad \hbox{for}\quad r>d
\end{eqnarray}
We can express Eq.(12) and the closure (20) in the variational form
\begin{eqnarray}
\frac{\delta I_{RISM}}{\delta c_{es}} = 0
\end{eqnarray}
where
\begin{eqnarray}
I_{RISM} = \rho_s\hat c_{es}(0) + 
{1\over 2}\int \frac{d^Dk}{(2\pi)^D}\hat c_{es}^2(k)\hat\chi_{ss}(k)
\hat\omega(k)
\end{eqnarray}
where $\hat c_{es}(0)$ is the k = 0 spatial Fourier transform of $c_{es}(r)$,
$D$ is the dimensionality of the space,  $d^Dk$ is the 
dimensionality dependence volume element,  $u$ labels the beads 
in the polymer ring $(0\leq u \leq \beta\hbar)$, $m$
is the bare electron mass, and $v(k)= -\hat c_{es}^2(k)\hat\chi_{ss}(k)$ 
is the Fourier transform of the potential between beads, which is found in 
Eq.(10). The information about the
electron-solvent atom interation is contained in the closure relation.

In Eq.(11), $\omega$({\bf r}) is the intra polymer distribution
function
averaged over all beads of the ring polymer. In writing Eq.(11) it has
been assumed that for each polymer configuration, the solvent sees only
average polymer rather than individual beads. The intra polymer distribution
function $\omega(r,\tau)$ is determined in the {\it polaron approximation} 
\cite{ref1},$\omega(k,u)$ is the Fourier transform of $\omega(r,u)$ and is given by
\begin{eqnarray}
\hat\omega (k,u) = \exp [-k^2R^2(u)/2D]. 
\end{eqnarray}
  Eq.(22) is solved self-consistently \cite{ref30} for a given
solvent and model
potential representing the electron-fluid particle interaction. This
solution gives information about $v({\bf r})$, $\omega({\bf r})$, $R(u)$,
$\gamma_n$ and $g$(r) [ = 1+$h$(r)]. Note that the quantity 
{\it R}(u) is the root mean square (RMS) value of the displacement between 
two points on the 
electron path separated by a time increment $0 \leq u \leq \beta\hbar$. The
characteristic size or breadth of the polymer is measured by
$R(\beta\hbar/2)$. This is a measure of the spread of the wave packet
associated with the particle. Since in the CSR theory a periodic boundary
condition , ${\bf r}(0) = {\bf r}(\beta\hbar)$, has been imposed on the
path of the electron, $R(u)$ is found to be symmetric about $u = 
{1\over2}\beta\hbar$, {\it i.e.} it starts from zero at u = 0, attain a 
maximum
value at $u = {1\over 2}\beta\hbar$ and decreases for 
$u > {\beta\hbar\over2}$ reaching zero at $u = \beta\hbar$. $g(r)$ 
gives information about
the average packing of solvent particles around the electron. The
variational parameter $\gamma_n$ measures the strength of the electron
fluid coupling. Quantities such as average kinetic energy, potential energy 
and effective mass etc., can be expressed in terms of $\gamma_n$ 
\cite{ref9,ref24,ref27} as 
\begin{eqnarray}
\langle K.E. \rangle = {D \over 2}k_BT\bigg[1 + 
{\gamma_n \over \beta m \Omega_n^2 + \gamma_n}\bigg]. 
\end{eqnarray}
\begin{eqnarray}
\langle P.E.\rangle = \rho_s \int d{\bf r} u_{es}(|{\bf r}|) 
g(|{\bf r}|). 
\end{eqnarray}
\begin{eqnarray}
{m\over m^\ast} = 24\sum_{n\ge 0}^\infty 
(4\pi^2 n^2 + \gamma_n \lambda_e^2)^{-1}. 
\end{eqnarray}
In this equation $m^\ast$ is the effective mass of the solvated electron.
  As mentioned earlier, we need two input for this theory. One is 
electron-solvent and another is the density-density correlation function of 
solvent which is related to the structure factor of the solvent. For 
$D$-dimensional hard sphere solvent under cocsideration the Percus-Yevic (PY) 
equation can be solved analytically for $D$ = 1 \cite{ref31} and 
for $D$ = 3 \cite{ref32}. For $D=2$ excellent results of thermodynamic and 
structural properties have been obtained by Baus and Colot \cite{ref33}. In the 
above $\hat\omega$, $\hat c_{es}$, and $\hat\chi$, are the spatial 
Fourier transform of $\omega$, $c_{es}$, and $\chi$. 

\section{Results and Discussions}
  In presenting our results we mainly focus on the imaginary time 
correlation function
\begin{eqnarray}
 R^2({1 \over 2}\beta \hbar) = \langle |{\bf r}({1 \over 2}\beta \hbar) - 
{\bf r}(0)|^2 \rangle  = 4D \sum_{n \ge 1} A_n (1 -\cos \pi n)
\end{eqnarray}
and the electron-solvent radial distribution function. Note that 
$R({1\over 2} \beta \hbar)$ gives a measure of the physical size of the 
electron 
chain or the spread of the wave packet associated with the electron. For a 
free 
particle,
\begin{eqnarray}
R({1\over 2} \beta \hbar) = \sqrt{\mathstrut D/4}\phantom{a} \lambda_e.
\end{eqnarray}

  In Fig.1 we plot the reduced correlation length, {\it S} 
$\equiv R({1\over 2}\beta\hbar)/
\sqrt{D/4}\phantom{a} \lambda_e$ which is the dispersion of the wavepacket 
associated with the solvated electron relative
to the free particle in one, two and three dimension  as a function of 
density for 
$\lambda_e = 15\sigma$, $\alpha$ = $\sigma^{-1}$ and $d/\sigma$ = 0.29 
for several values of the  attractive interaction, $\beta\epsilon$.  
From these figures we find that when the electron-solvent interaction is 
solely repulsive ($\beta\epsilon = 0.0$), the electron is always gets trapped 
inside a solvent cage as solvent density is increased in one and two 
dimensions. In three dimension [see Fig. (1C)] the repulsive interaction 
is not strong enough (because $d/\sigma = 0.29$) to localize the electron 
due to ordered structure formed when the solvent density is high.
When the attractive interaction between electron and solvent atom is large 
($\beta\epsilon \simeq 100$), the reduced correlation length of the electron 
is very small at very 
low solvent density and it stays almost constant upto $\rho^\ast \sim 0.5$.
This behavior strongly indicates the electron is localized to a single atom 
irrespective of space dimensionality. 
As the space dimensionality increases, the range of the constant value of the 
solvent density decreases for example for $D$=1, {\it S} is constant upto 
$ \rho^\ast \simeq 0.6$. One interesting 
feature we have noted from Figs. 1A
, 1B, and 1C is that as  
the attractive interaction increases, the reduced correlation length of the 
electron decreases as we increase the solvent density in the low-density 
regime, 
but the correlation turns upwardto have a maximum in the middle density 
regime, and 
finally the electron is trapped in the high density regime. If we carefully 
look at Figs.1, we observe that as we increase the space dimensionality, 
we find these effect are pronounsed for large attractive interaction. The 
similar type of 
behavior has been observed for the other values of $d/\sigma$. When $d/\sigma$ 
is less than $0.29$ we found the reduced correlation length of the electron 
decreases as we increase the attractive interaction {\it i.e.} 
$\beta \epsilon$ 
and is self trapped at lower density. Opposite trend is found for increasing 
$d/\sigma$.  

  From Figs. 1[A], 1[B] and 1[C] it is clear that at the begining when
we start increasing the attractive interaction ($\beta\epsilon \simeq 20$) 
the wave packet associated with the electron increases then further increase 
of electron solvent attractive interaction i.e. when 
$ \beta\epsilon > 50.0$ the electron is localized on a single solvent atom. 
The reason behind the delocalized state at small vales of attractive 
interaction ($\beta\epsilon$) is the cancellation between the repulsive and 
attractive electron-solvent interaction.  
  In Fig. 2[A] to Fig.2[F] we plot the electron-solvent radial 
distribution function for one, two and three dimension at various 
solvent density and attractive iteraction ($\beta \epsilon$). 
Figs. 2 can be explained in consistent manner with the same physical picture 
given for Figs. [1]. When the attractive interaction is weak, the electron 
pushes the solvent atoms makes space to self-trap.
When the 
attractive interaction 
is very large ($\beta \epsilon \ge 100.0$) the electron is trapped 
on a single atom 
irrespective of space dimensionality, as evidenced by a large peak in $g(r)$ 
[see Figs. 2[F].

In Fig. 3[A], 3[B], and 3[C] we plot the reduced correlation length 
as a function of $\lambda_e/\sigma$ 
at $\beta \epsilon 
= 0$ for $D$=1, 2, and 3 respectively. 
In one and two dimension ( Fig.3[A] and Fig.3[B] ) we observed as we 
increase $\lambda_e$ 
the electron is trapped in the {\it solvent cage}. If we compare Fig. 3A 
with Fig. 3B as the space dimensionality increases, the 
electron is strongly 
trapped in the {\it solvent cage}. One interesting feature we found 
in Fig. 3[C] 
at $\rho^\ast \sim 0.7$ there is sharp transition from delocalized to 
self-trapped state. In less than  two dimension when there is no attractive 
interaction, the electron will always be in a localized state\cite{ref16} but in 
three dimension there is always a tendendency the electron will be in 
delocalized 
state.  On the other hand, as the temperature is 
decreased, the electron has tendency  to be in the localized state. This 
competition between vaious length scales probably make the sharp transition 
around $\lambda_e \sim 16 \sigma$.

  In Fig. [4A], [4B] and [4C] we plot the reduced correlation length 
as a function of density ($\rho^\ast$) for various 
values of $d/\sigma$. We found for low $d/\sigma$ (for example $d/\sigma$ 
= 0.15) 
the attractive interaction dominates and the electron is self-trapped 
in the low density regime. On the other hand, when $d/\sigma$ is large the 
self-trapping of the electron is dominated by repulsive interaction or 
self-trapping is by {\it cage effect}. In the case of electron in 
helium and xenon, 
the self-trapping of electron in helium and delocalized state of the electron 
in xenon has been explained on the basis of 
$d/\sigma$ value \cite{ref25} when elecron-solvent interaction is repulsive 
{\it i.e.} $\beta\epsilon = 0$.  


\section{Concluding Remarks}
The CSR theory for the excess electrons in simple fluid ( consisting of 
spherical atoms )  has been studied in one, two, and three dimension. The 
detailed study led us to the following conclusions :  

\begin{enumerate}
\item The reduced correlation length, {\it S}, is 
very sensitive to the nature of the electron-fluid atom interaction, thermal 
wavelength of the excess electron ( $\lambda_e$ ) and a length associated 
with the mean volume occupied by each fluid atom which is related to 
$\rho^{\ast ^{-{1\over D}}}$, where $\rho^\ast = \rho\sigma^D$, $\rho$ 
being the 
number density of the fluid atoms and $\sigma$ is the effective diameter of a 
fluid atom. The behavior is interpreted in terms of an interplay among the 
length scales noted above. When the electron-solvent attractive interaction is 
very large, the density dependence of the size of the electron polymer 
relative to the free particle is dominated essentially by the pair 
attractive interaction, and the electron is trapped in a single solvent atom 
irrespective of the dimensionality ($D$). On the contrary, if the attractive 
interaction is absent, the electron is trapped in a cage formed by the 
solvent atoms at higher density. When the attractive interaction is low 
($\beta\epsilon \sim 10.0$) due to cancellation between repulsive and 
attractive interaction, the electron is delocalized irrespective of 
the dimensionality. 

\item In the one dimensional case the reduced correlation length is almost 
independent of $d/\sigma$ when 
electron-solvent attractive interaction is absent. 
On the other hand, in two and three dimensions, $S$, the reduced correlation 
length of the electron is sensitive on the value of the $d/\sigma$. 

\item In three dimension we found when $d/\sigma = 0.29$ and 
$\beta \epsilon = 0.0$, the temperature dependence of the reduced correlation length, $S$, shows a sharp transition (metal-insulator type) around 
$\lambda_e/\sigma \sim 15$ at $\rho^\ast \sim 0.7$. 

\item The long range electron-solvent attraction dominates at low density, 
while hard core repulsion dominates at high densities. When both 
interactions are present, they can counterbalance each other. Consequently, 
at some intermediate density, the effective electron-solvent 
interaction can be quite small, resulting in the electron delocalization.  
\end{enumerate}

  In the present work we have explored the electron self-trapping in 
simple liquid for various space dimension ($D$= 1,2, and 3) which have same 
model potential (solvent-solvent interaction potential and electron-solvent 
interation potential) in every space dimension. In every space dimension (i.e. 
$D$=1, 2, and 3) we have in the model system considered here there is 
repulsive as well as attractive electron solvent interation potential which is 
not considered in Refs. 25, 27, and 28. We have presented the dimensionality 
dependent result for self trapping behavior of electron in simple liquid.

\begin{acknowledgments}
This work was supported in part by National Science Foundation, the Robert 
Welch Foundation,  and the Institute 
for Molecular Science, (Japan).
\end{acknowledgments}

\appendix
\section{ $D$-space dimensional integration }

 In $D$-space we have for any function f({\bf x}) depending only on
the distance $x = |{\bf x}|$,
$$ \int d{\bf x} f(x) = S_D \int_0^\infty dx x^{D-1} f(x),\eqno (A.1)$$
  or on $x$ and one integration angle $\theta$,
$$\int d{\bf x} f(x,\theta) = S_{D-1}\int_0^\infty dx x^{D-1} \int d\theta 
\sin^{D-2}\theta f(x,\theta),\eqno (A.2)$$
  where $ S_D = D V_D$ is the surface area of the unit sphere of 
volume $V_D = \pi^{D/2}/\Gamma(1 + D/2)$,where $\Gamma(1 +z) = z!$ is the 
$\Gamma$ function.

\begin{figure}[h]
\includegraphics{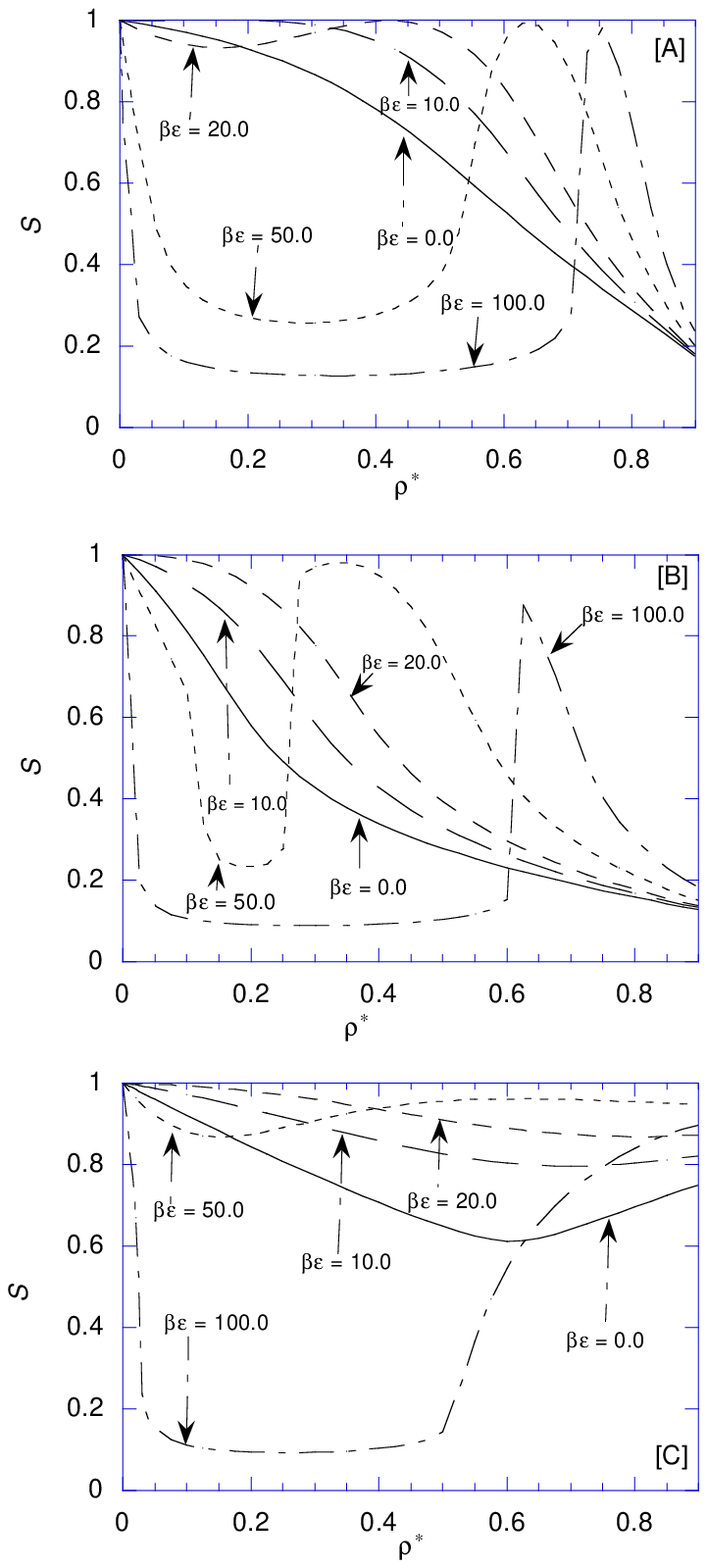}
\caption{Density ( $\rho^\ast = \rho_s \sigma^D$ ) dependence of reduced 
correlation length $S$ for various values of attractive interaction 
( $\beta\epsilon$ ) with $\lambda_e = 15\sigma$, $\alpha$ = 
$\sigma^{-1}$ and $d/\sigma$ = 0.29 (A) for one dimension ( $D$ = 1 ), 
(B) for two 
dimension ( $D$ = 2 ), and (C) for three dimension ( $D$ = 3 ).}
\end{figure}

\begin{figure}[h]
\includegraphics{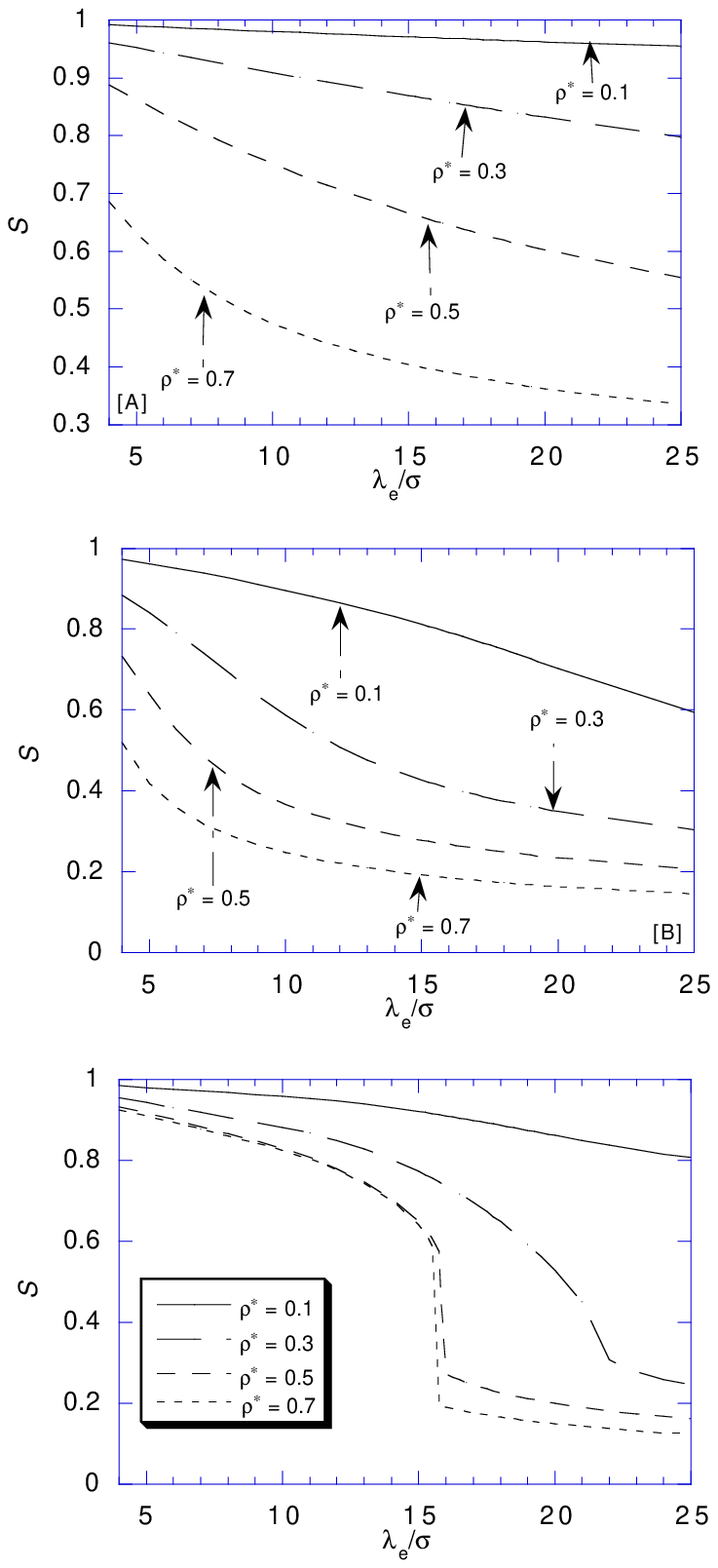}

\caption{Electron-solvent radial distribution function g(r) for an electron in 
D-dimensional fluid at various density ( $\rho^\ast$ ) with $d/\sigma = 0.29$, 
$\lambda_e = 15\sigma$, and $\alpha = \sigma^{-1}$ for various values of 
attractive interaction ($\beta\epsilon$) and various dimensionalities. }
\end{figure}

\begin{figure}[h]
\includegraphics{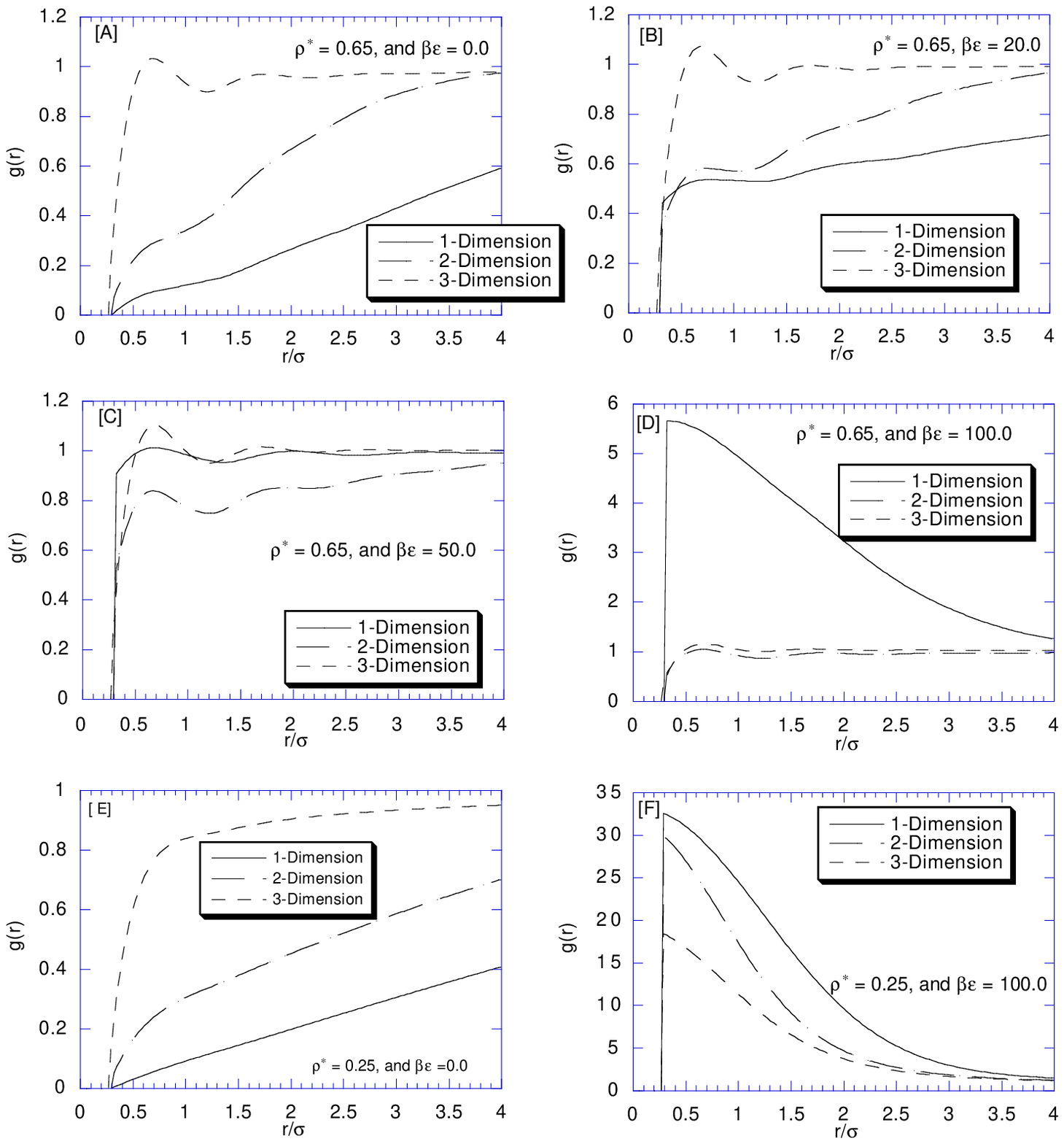}

\caption{Temperature dependence of the reduced correlation length ,$S$, relative 
to the free paricle value at 
$\alpha = \sigma^{-1}$ and $d/\sigma$ = 0.29 for various values of density 
[A] for one dimension (upper panel) ( $D$ = 1 ), [B] for two 
dimension ( $D$ = 2 ) (middle panel), and [C] for three dimension ( $D$ = 3 )
(lower panel). }
\end{figure}

\begin{figure}[h]
\includegraphics{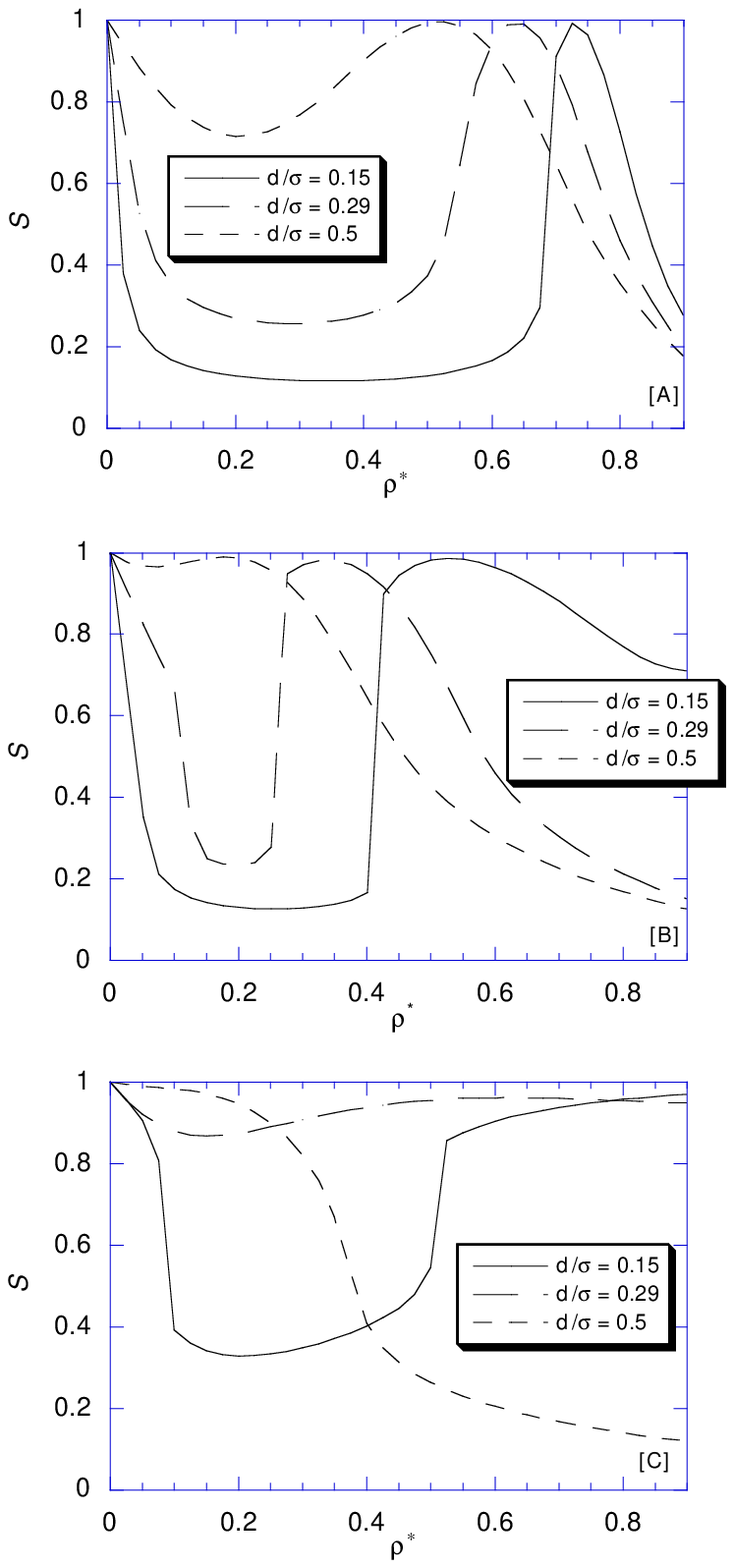}

\caption{Density dependence of the reduced correlation length at 
$\beta\epsilon = 50.0$,  $\lambda_e = 15\sigma$, and 
$\alpha = \sigma^{-1}$for various values of $d/\sigma$ 
[A] for one dimension [B] for two dimensions [C] for three dimensions. }
\end{figure}

%

\end{document}